\documentclass[twocolumn]{article}

\usepackage[utf8]{inputenc}
\usepackage{siunitx}
\usepackage{array,url,kantlipsum}
\usepackage[pdftex]{graphicx}
\newcommand{\Mark}[1]{\textsuperscript{#1}}
\usepackage{hyperref}
\usepackage[super,sort&compress,comma]{natbib} 
\usepackage[left=1.7cm, right=1.7cm, top=1.785cm, bottom=2.0cm]{geometry}
\usepackage[format=plain,justification=justified,singlelinecheck=false,font={stretch=1.125,small,sf},labelfont=bf]{caption} 
\usepackage{balance}
\usepackage{textcomp}
\usepackage{xcolor}
\usepackage{lmodern}

\begin{document}
\sffamily
\twocolumn[{%
 \LARGE Droplet-confined electroplating for nanoscale additive manufacturing: current control of the initial stages of the growth of copper nanowires \\[1.em]

 
 \large Mirco Nydegger\Mark{1}
       and Ralph Spolenak\Mark{1,}$^{*}$
       \\[1em]

 \normalsize
 
\Mark{1}\small{Laboratory for Nanometallurgy, Department of Materials, ETH Zürich, Vladimir-Prelog-Weg 1-5/10, Zürich 8093, Switzerland}\\

  $^{*}$To whom correspondence should be addressed; e-mail: ralph.spolenak@mat.ethz.ch\\[0.5em]

\textbf{Droplet-confined electrodeposition enables a precise deposition of three dimensional, nanoscopic and high purity metal structures. It aspires to fabricate intricate microelectronic devices, metamaterials, plasmonic structures and functionalized surfaces. Yet, a major handicap of droplet-confined electrodeposition is the current lack of control over the process, which is owed to its dynamic nature and the nanoscopic size of the involved droplets. The deposition current offers itself as an obvious and real-time window into the deposition. Therefore, the current during droplet-confined deposition is analysed. Nucleation and growth dynamics are evaluated systematically. Our results indicate different deposition regimes and link current to both volume and morphology of deposited copper. This allows for optimized electroplating strategies and to calibrate the slicing algorithms necessary for a controlled deposition of 3D structures. The potential of selecting appropriate solvents further readies this novel technique for the reliable deposition of functional structures with submicron resolution.}\\[1em]
 
}]

\section{Introduction}

\begin{figure*}[!ht]
 \centering
 \includegraphics[width=\textwidth]{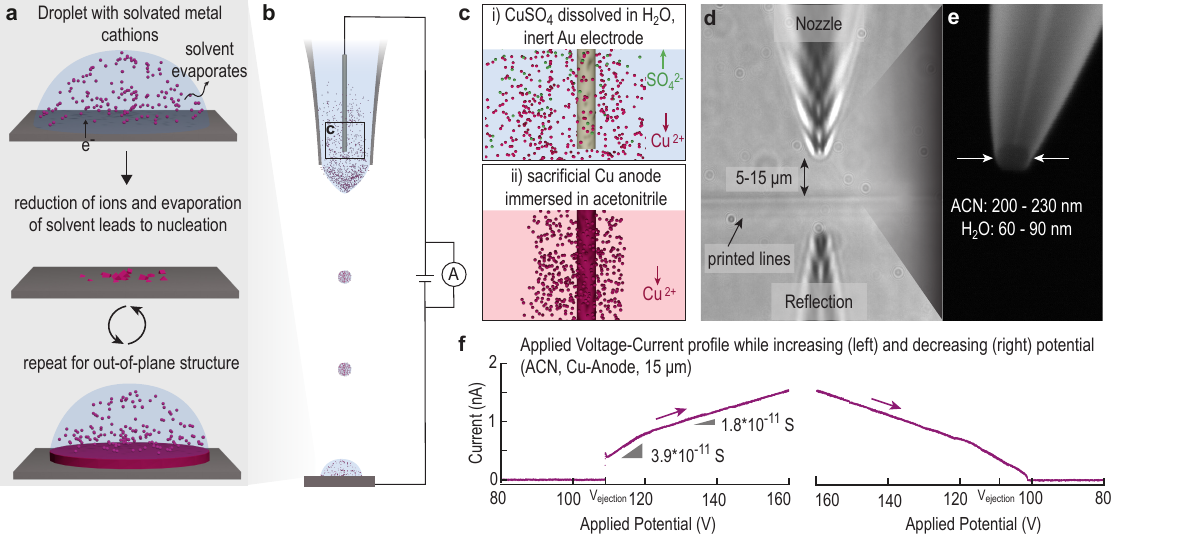}
 \caption{\textbf{Droplet-confined electroplating utilizing electrohydrodynamic redox 3D printing} \textbf{(a)} We spatially confine electroplating into nanoscopic solvent droplets. It is important to note that these droplets contain only metal-cations and solvent molecules, as we do not observe the elements comprising the counterions in EDX spectra or mass spectra under normal deposition conditions (shown in previous studies\cite{Reiser2019Multi-metalScale, Nydegger2022AdditiveStructures}). On the substrate, the metal ions are reduced while the solvent evaporates. A repetition of this cycle leads to the formation of an out-of-plane metal nanostructure. \textbf{(b)} The solvent droplets with solvated metal ions can be generated and deposited using electrohydrodynamic (EHD) ejection, which has been described in detail previously\cite{Park2007High-resolutionPrinting, Lee2013OptimizationPrinting, Reiser2019Multi-metalScale}. Here, a solvent filled quartz capillary is used as a nozzle and gold coated wafers are used as substrates. \textbf{(c)} For the present study, we utilize and compare two different ion sources and solvents that have been successfully used in previous work: dissolved metal salts in low concentrations (1~mM CuSO\textsubscript{4} in H\textsubscript{2}O)\cite{Porenta2023Micron-scalePrinting} and sacrificial metal anodes in acetonitrile (ACN)\cite{Reiser2019Multi-metalScale, Menetrey2022TargetedPrinting}. \textbf{(d)} Camera image showing the printing process: the nozzle (top) is in close proximity (5\textendash  15 \textmu m) of the substrate. At the bottom, the reflection of the nozzle is visible, while printed walls appear as diffuse lines. \textbf{(e)} Scanning electron micrographs verify the orifice of the nozzle to be between 200 to 230~nm for acetonitrile and 60 to 90~nm for dilute metal salt solutions in H\textsubscript{2}O. \textbf{(f)} Voltammetry of a deposition with ACN as the solvent and a copper anode (scan speed 50~mV/s). We can distinguish different regimes in the I-V (current-potential) curve when increasing the potential (left graph). At $V < V\textsubscript{ejection}$ no ejection is observed and therefor a current of zero is measured. At $ V = V\textsubscript{ejection}$ ejection of ion-loaded droplets starts. The start is not a gradual increase in the current but rather a sudden change to around 0.5~nA. Above a current of around 0.8\textendash 1~nA the slope of the I-V curve changes: an decrease in the conductivity is observed (from 3.9*10\textsuperscript{-11}~S to 1.8*10\textsuperscript{-11}~S). When the potential is decreased (right graph) a similar curve is observed with a change in the slope at the same current. However, the curve is continuously until a current of 0 is measured. Therefore low deposition currents (I $<$ 0.5~nA) can be accessed by starting the deposition at a high potential and then reducing the potential again.}
  \label{fig_technique}
\end{figure*}

Intricate, nanostructured metals have exciting applications in integrated circuits and MEMS devices\cite{Hu2010Meniscus-ConfinedBonds, Ulisse2022ATubes}, nanorobots\cite{Moo2017Nano/MicrorobotsElectrochemistry} and plasmonic structures\cite{Menetrey2024NanodropletPrinting}. Small-length scale additive manufacturing (AM) aspires to provide access to such three dimensional structures at the nanoscale without the geometrical constraints of two-dimensional lithography. Small-length scale AM techniques based on localized electrodeposition\cite{Braun2016TheManufacturing, Hengsteler2023BeginnersManufacturing} are especially well-suited for the deposition of device-grade metals, due to a high purity and a dense microstructure\cite{Hirt2017AdditiveScale,Reiser2020MetalsProperties} of plated metals. Such plated metals often exhibit excellent mechanical and electrical properties\cite{Reiser2020MetalsProperties}. Therefore, small-length scale electrochemical AM has drawn significant attention from both fundamental and industrial research. Such small-length scale electrochemical AM techniques rely on a localization of one of the fundamental steps of electroplating: a localized supply of both electrons and cations or a spatial confinement of the electrolyte. 

\begin{figure*}[!ht]
 \centering
 \includegraphics{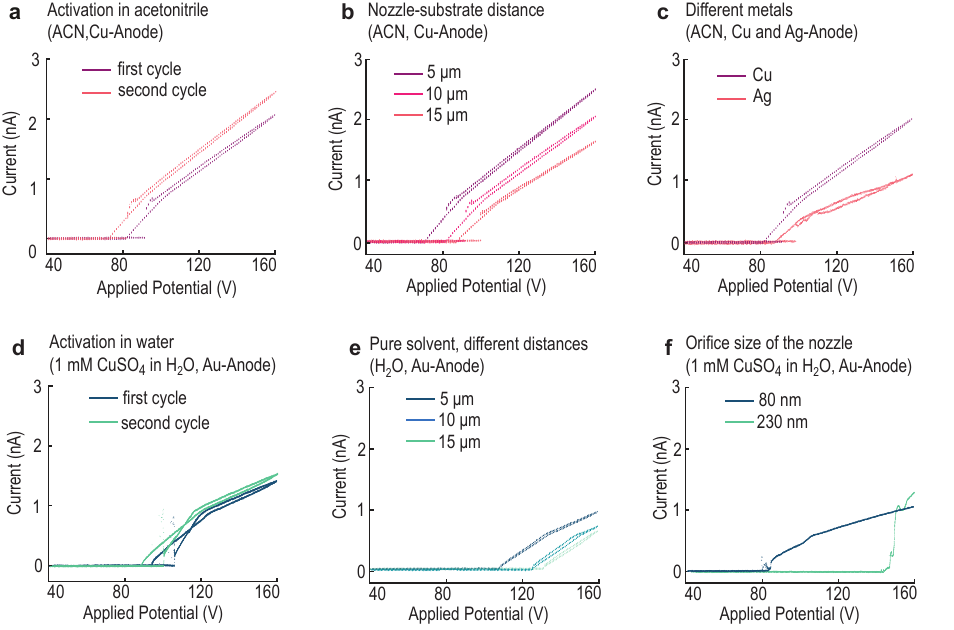}
 \caption{\textbf{: Voltammetry measurements of EHD-RP.} \textbf{(a)} The measured current in EHD-RP changes with the elapsed time of the deposition. Two cycles from 0 to 160~V and back with the same nozzle and electrode do not show the same current values. In the second cycle the curve is shifted to higher currents for the same potential, indicating an activation of the surface. Note that the current at which the slope changes stays constant (Scan speed 500~mV/s). \textbf{(b)} The nozzle-to-substrate distance has a strong influence on the current. A smaller distance yields a higher current for the same potential and lowers the necessary potential to start ejection. Furthermore, the current at which the resistance increases is higher at smaller distances. \textbf{(c)} The element that is used as a sacrifical anode has also a large influence on the observed current. It is to know that copper oxidizes more readily compared to a silver anode, which leads to a higher current for the same potential. \textbf{(d)} The current-curve of copper sulfate dissolved in water exhibits a similar change in the slope around a current of 0.8\textendash 1~nA. The onset of the ejection is unstable with currents observed varying between 0 and 1~nA. A second cycle shifts the curve towards higher current similarly as observed for ACN. (The scan speed is 50~mV/s to illustrate the start of the ejection) \textbf{(e)} Ejection of copper-free water is possible, but requires high potentials to start the ejection (presumably of hydroxonium loaded droplets, that is reduced to hydrogen on the cathode) unless the nozzle is very close to the substrate. Without copper ions present only low currents are achieved and no unstable current was observed upon start of the ejection (\textit{i.e.} fluctuating current). Increasing and decreasing the potential yields parallel current-curves. \textbf{(f)} The size of the orifice has a large influence on the ejection when using water-based solvents: small orifices (60\textendash 90~nm) enable an ejection starting from a similar potential (ca. 80~V) as in ACN. In contrast, 230~nm nozzles show a very different current-potential behaviour with a steep increase in the current after ejection start. This steep increase in current will render the deposition difficult to control, therefore narrower orifices are used to deposit aqueous electrolytes.}
  \label{fig_CV}
\end{figure*}

\textit{Localized supply of electrons:} The localized supply of electrons for reduction can be achieved by using focused electron beams (such as in electron microscope)\cite{Esfandiarpour2017FocusedSolutions} or by utilizing sharp electrodes (such as sharp scanning tunneling microscope tips)\cite{Reiser2022NanoscaleSTM}. Here, the short mean free path of electrons in solvents will spatially constrict the deposition\cite{Meesungnoen2002}. Despite the potential for layer-by-layer deposition, the fabrication of arbitrarily shaped, three dimensional structures has not yet been demonstrated to our knowledge. 

\textit{Localized supply of cations}: Cations can be provided locally by immersing hollow AFM tips (refered to as FluidFM) \cite{Schurch2023DirectElectrodeposition} or quartz nozzles in a supporting electrolyte, in a setup similar to scanning ion electrochemical microscopy\cite{Momotenko2016Write-ReadNanopipette}. This allows to produce high-fidelity structures from a variety of metals\cite{vanNisselroy2022ElectrochemicalFluidFM}. 

\textit{Confinement of the electrolyte}: A more common approach is to utilize a mask so that only parts of the conductive substrate are exposed to the electrolyte. This will limit the deposition into the exposed area and leave an out-of-plane structure when the mask is removed. This approach is often used to synthesize nanowires\cite{Moo2017Nano/MicrorobotsElectrochemistry}, but also more intricate structures \cite{Braun2016TheManufacturing, Gunderson2021NanoscalePyrolysis}. Electroplating confined by a meniscus between the substrate and a nozzle offers a flexible mask and therefore a higher flexibility for intricate designs. This technique further allows for the fabrication of high resolution structures ($<$40nm)\cite{Hengsteler2021BringingNanoscale}. 

A fundamental limitation shared by all ECAM techniques is the low deposition speed compared to other small-length scale AM techniques due to the necessary migration of cations. The migration of cations can be sped up by a forced mass-transfer by ejection of ion-loaded solvent droplets in an electric field. The forced mass-transfer is then followed by a spatially confined electroplating in droplets\cite{Reiser2019Multi-metalScale}. This technique, known as electrohydrodynamic redox 3D printing (EHD-RP), allows for a fabrication of complex structures with speed of around 100 voxels/s, the voxel size being around 60~nm\cite{Menetrey2024NanodropletPrinting}. The voxel size directly depends on the size of the solvent droplet, which is generated from a quartz capillary (Fig. \ref{fig_technique} a-e). Therefore, EHD-RP combines high deposition speed typical for material-transfer AM techniques with the dense microstructure of electroplated metals. Combining this technique with an electron microscope allows to electroplate metals directly onto insulating substrates\cite{Nydegger2024DirectSize}.

It is important to note that droplet confinement is conceptually not merely an extension of standard bulk electrochemical processes, but works in a different regime of electroplating with different governing mechanisms. In particular, the transient nature of the droplet as well as the absence of a liquid connection between the electrodes are unique features of EHD-RP. This allows, for example, direct deposition of alloys with a controlled chemical composition \cite{Porenta2023Micron-scalePrinting} without the need for additives. Furthermore, the deposited metal can be switched in a voxel-by-voxel fashion on-the-fly\cite{Reiser2019Multi-metalScale} to create chemically architected materials. Yet, several fundamental questions still need to be addressed to fully leverage this novel technique routinely in nanofabrication. Specifically, to ready this technique for the fabrication of complex structures, a direct feedback mechanism for the printed volume has to be established. This is fundamental to optimize the discretization of a planned design into individual voxels to be printed (\textit{slicing}) and is therefore crucial for the reliable fabrication of intricate nanostructures. Further, the nucleation and current-microstructure relationships in this novel regime of electroplating have only partially been described\cite{Menetrey2022TargetedPrinting}, despite their importance for the fabrication of metals with dense microstructures. Further, an accurate current control promises to optimize the resolution of droplet-confined electroplating.

In this work, we first investigate the deposition current by mapping the relationship between applied potential and measured current in EHD-RP for different solvent systems and different distances between nozzle and substrate. We then investigate the nucleation in droplet-confined electroplating at different currents. Lastly, we establish a relationship between deposition current and the deposited voxels. We limit the number of voxels that are investigated to 4 to limit the influence of other effects such as field focusing\cite{Menetrey2024NanodropletPrinting}.

\section{Results}
\begin{figure*}[!ht]
 \centering 
 \includegraphics[width=\textwidth]{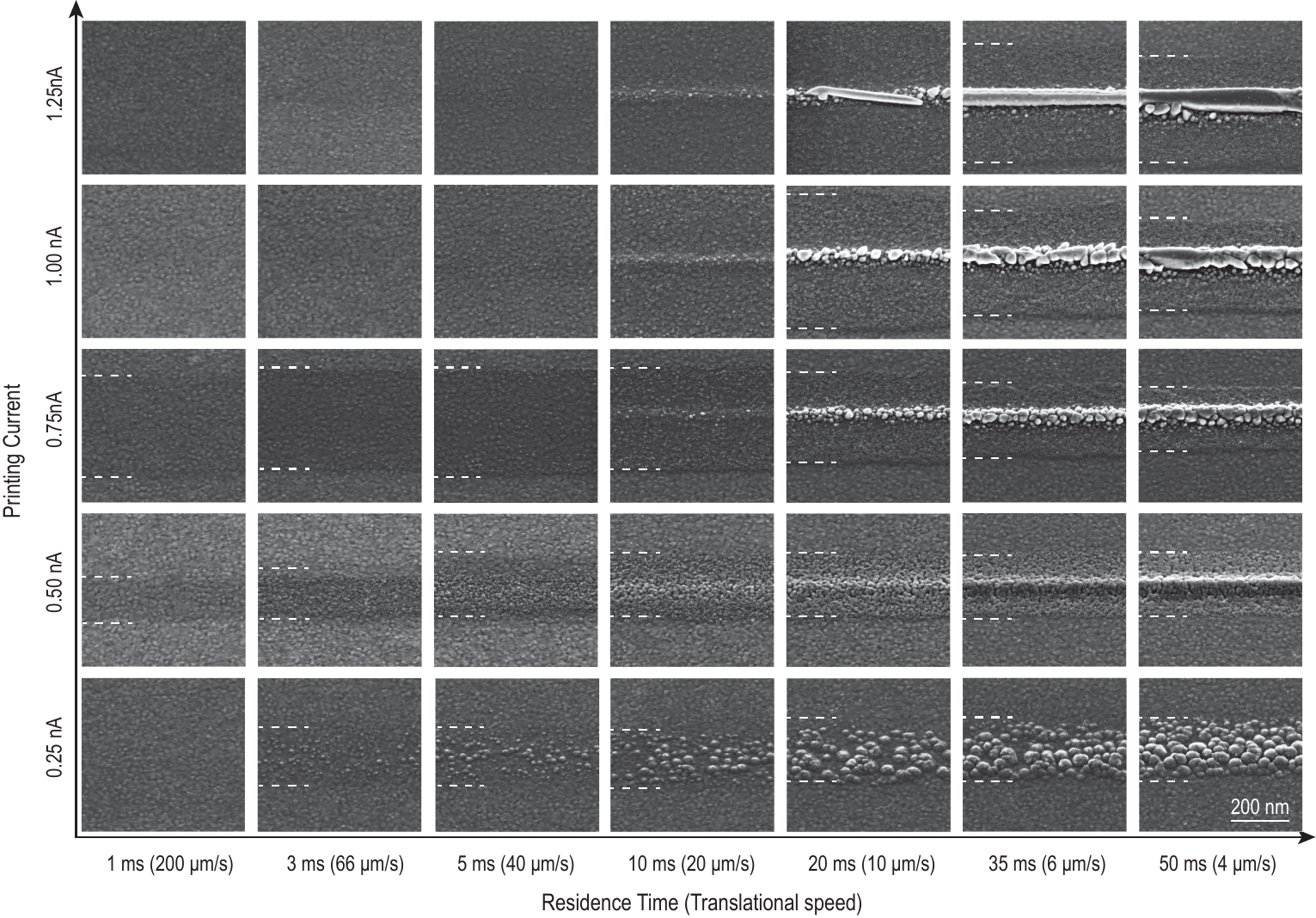}
 \caption{\textbf{Nucleation of Cu on Au, deposited in acetonitrile.} To arrest the deposition at short time steps, lines were printed with different speeds (nozzle-to-substrate distance of 10~\textmu m). We approximate the time stamps for the deposition by using the residence time, which is derived by dividing the orifice size of the nozzle by the line speed. The dashed white lines indicate the area, in which nuclei can be spotted. For 1 and 1.25~nA and residence times $<$ 20~ms this area is larger than the area shown in the micrographs, while for 0.25~nA and 1~ms residence time probably not enough material was deposited to be visible. The substrate surface consists of polycrystalline Au which is visible in the area outside the dashed white lines. A current of 0.25~nA leads to patchy, non-connected deposits of copper. These patches then increase in size with longer residence time, but the area between the dashed lines remains constant over longer residence times. At 0.5~nA finer grains are deposited that form a continuous, but rough structure at longer residence time. The area in which nuclei can be found also stays constant. For 0.75~nA and higher currents individual Cu islands can be seen. Progressively fewer, but larger grains are observed with increasing current until at 1.25~nA only a single large grain is observed. For residence times below 10~ms almost no nuclei can be observed for these currents. Also, the area, in which nuclei can be seen, decreases with increasing residence time (from left to right) for currents of 0.75~nA and higher. This reduction in area indicates more localization of the deposition with longer residence times.}
  \label{fig_NucleationACN}
\end{figure*} 

\begin{figure*}[!ht]
 \includegraphics[width=\textwidth]{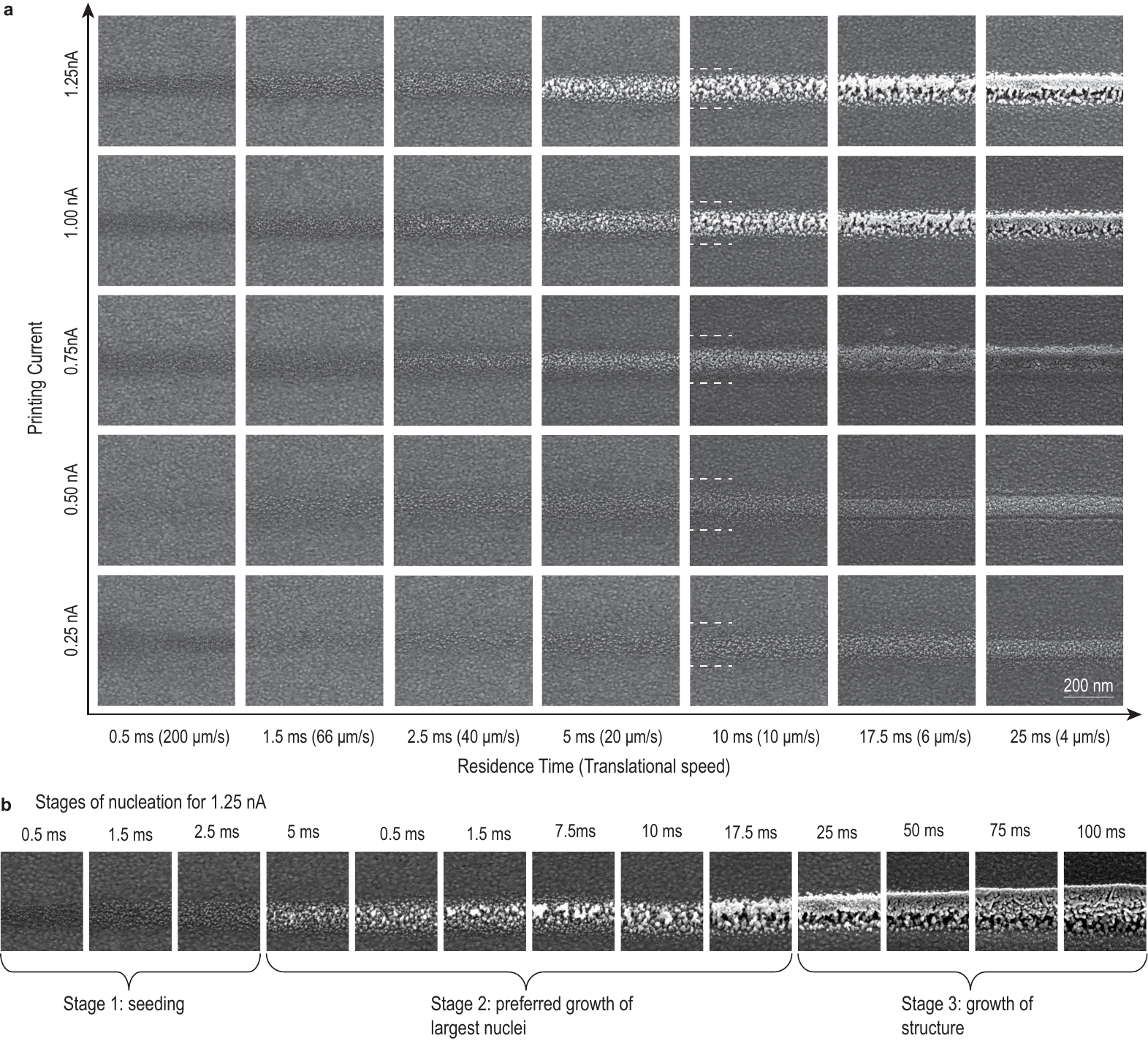}
 \caption{\textbf{Nucleation of Cu on Au, deposited from 1 mM CuSO\textsubscript{4} in H\textsubscript{2}O.} \textbf{(a)} Similarly to ACN, Cu lines were printed using a 1 mM CuSO\textsubscript{4} solution (nozzle-to-substrate distance of 7.5~\textmu m). The nucleation is shown at shorter residence times (due to the narrower orifice sizes of the nozzles but the same utilized velocity). The dashed white lines indicate again the area in which nucleation is observed. For 0.25 and 0.5~nA only small grains are observed, but no patches as in ACN. At 0.75~nA the grains coalesce into a continuous but flat structure. For 1 and 1.25~nA lines consisting of fine grains are observed and not the large crystals that are observed in ACN. At 1.25~nA and 25~ms residence time a horizontal pore is observed at the interface between substrate and printed Cu line. In general, it appears that less material is deposited (this will be shown in more detail in Fig. \ref{fig_crosssection_CuSO4}). Furthermore, the area in which nuclei are observed stays constant over different currents and residence times. \textbf{(b)} Based on the observed nucleation behaviour we can distinguish different stages of the nucleation for 1.25~nA: First, with short residence times, a number of similar sized nuclei are formed (Stage 1). Then, from 5 ms residence time onwards, some nuclei grow preferably. This leads to many small and few large grains after 10 ms (Stage 2). After around 20 ms the growth behaviour changes again: here, the large nuclei start to coalesce (Stage 3). This forms a capping layer on top of the porous first layer. This capping layer then grows further with longer residence times.}
  \label{fig_NucleationCuSO4}
\end{figure*} 

\subsection{Voltammogram of electrohydrodynamic redox printing}
Deposition in EHD-RP is assumed to be confined into transient solvent-droplets without a liquid connection between working- and counterelectrode. This unique working principle is reflected by a distinctive shape of the voltammograms (Plots of applied potential versus the measured current) of EHD-RP (Fig. \ref{fig_technique}f and Fig. \ref{fig_CV}). The voltammograms are measured by applying a constant potential between the immersed electrode (which acts as the anode in the setup) and the substrate (the cathode), while measuring the current for 1~s.The applied potential is then increased by $\Delta V$ (Fig. \ref{fig_technique}b). The applied potential is cycled between 0 and 160~V with a $\Delta V$ of 0.5~V (resulting in a scan speed of 500~mV/s, unless stated otherwise). For these experiments, the printed geometry consisted of parallel lines while maintaining a substrate-nozzle distance of 10 \textmu m (or between 5\textendash 15 \textmu m when investigating the influence of the distance). To verify that the observed current-potential relationship is not an artefact of the used amperemeter or setup, the anode was replaced with a 100~GOhm resistor (selected to match the current at 100~V) and connected to the substrate. With the wire replaced by a resistor, the current did not show such change in the slope (Supporting Information (SI) Fig. \ref{fig_supp_Current}a). Additionally, no current was detected when no solvent was added into the nozzle. We therefore conclude that the observed I-V curves are not due to artefacts in the setup.

The I-V curve in Figure \ref{fig_technique}f shows three distinct regimes characterized by different slopes of the curve. When a potential below the minimal ejection potential ($V < V\textsubscript{ejection}$) is applied, no current above noise level ($\pm 2*10^{-11}$~A) is measured. Here, the applied potential is not strong enough to start the hydrodynamic ejection of ion-loaded droplets. Note that in contrast to standard EHD printing no pressure is applied to the liquid in the nozzle \cite{Park2007High-resolutionPrinting}. The absence of a detectable current above noise-level includes the region around 0 ($0  \pm  4~V $) where a current from the dissolution of the Cu electrode could be expected, but is not observed (SI Fig. \ref{fig_supp_Current}b). The only current that is detectable for $V < V\textsubscript{ejection}$ is an induced current from changing the potential, since the experimental setup acts fundamentally as a capacitance. This induced current decays quickly if the $\Delta V$ is small ($\Delta V \leq 1 V$). When the potential is raised to above the minimal ejection potential $V\textsubscript{ejection}$ a sudden increase in current is observed. Usually, a potential of 75\textendash  90~V is necessary to start ejection. However, this minimal potential depends on further parameters, such the distance between nozzle and substrate, ion concentration and and the diameter of the nozzle (Fig. \ref{fig_CV}). For 5 \textmu m  distance and acetonitrile (ACN) as the solvent, a potential of $V\textsubscript{ejection} =$ 82~V is necessary, for 10 \textmu m $V\textsubscript{ejection} =$ 92~V, and for 15 \textmu m $V\textsubscript{ejection} =$ 100~V. Upon start of the ejection, the current raises instantaneous to ca. 0.5~nA for ACN. For water, the current fluctuates in the beginning between 0 and 1~nA (Fig. \ref{fig_CV}c), indicating an unstable ejection mode, but then stabilises at around 0.2~nA. A further increase of the potential leads to a linear increase of the current in both solvents (ACN: 3.9*10\textsuperscript{-11}~S, 1~mM CuSO\textsubscript{4}: 5*10\textsuperscript{-11}~S). Above a current of around 0.8\textendash 1~nA the conductivity decreases for both solvents (ACN: 1.8*10\textsuperscript{-11}~S, 1~mM CuSO\textsubscript{4}: 1.4*10\textsuperscript{-11}~S). A change in the conductivity is observed at the same current when the potential is reduced ($\Delta V < 0$, Fig. \ref{fig_technique}f). The main difference with $\Delta V < 0$ is that the ejection is continuously until a current of effectively 0 is reached. Therefore, deposition at low currents ($I <$ 0.3~nA for water and $I <$ 0.5 - 0.8~nA in ACN) can be achieved by first setting the potential so that a current of $<$ 0.5~nA is measured and then lowering the potential again. Droplets can also be ejected with a negative potentials as shown in SI Fig. \ref{fig_supp_Current}, but the ejection is not stable as indicated by the current curve. So far, no structures have been printed using negative applied potentials.

\begin{figure*}[!ht]
 \centering
 \includegraphics[scale = 0.85]{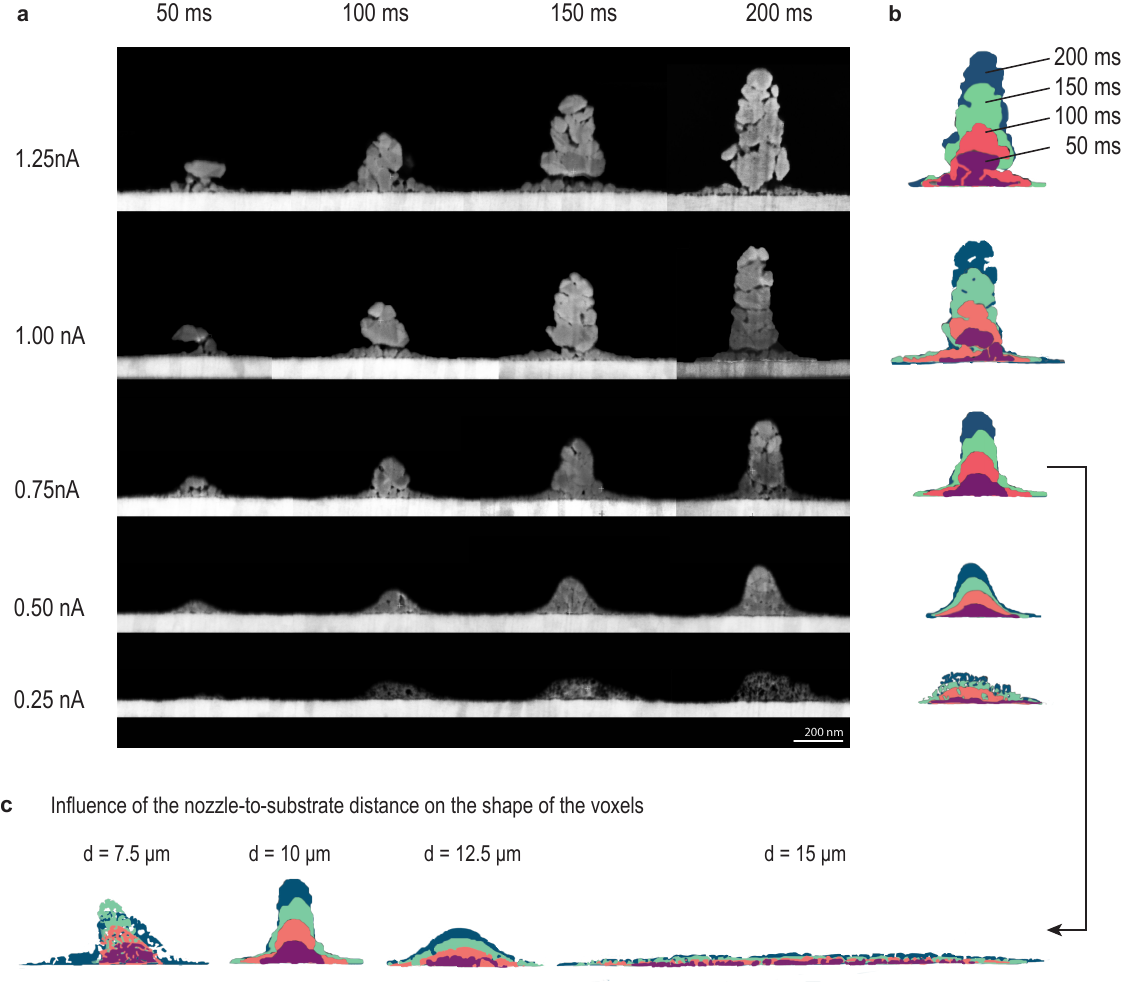}
 \caption{\textbf{Growth of Copper deposited in acetonitrile} \textbf{(a)} Cross-sectional micrographs of Cu lines printed in ACN with currents between 0.25\textendash 1.25~nA (nozzle-to-substrate distance of 10~\textmu m). The lines were printed with one or multiple overpasses of 50 ms each (so 150~ms refers to 3 overpasses), cut using a focused ion beam and imaged in SEM. Cu printed with 0.25~nA exhibits a very porous, low contrast structure resembling a network. With 0.5~nA a dense structure with a fine grain size is deposited. With higher currents structures exhibiting larger grains are deposited. This evolution of the grain structure is in line with the observations from Fig. \ref{fig_NucleationACN} and previous work (in a potential controlled mode)\cite{Menetrey2022TargetedPrinting}. A current above 0.75~nA increases the lateral thickness of the printed lines, which indicates that higher currents deposit more material but also show more porosity in between the large grains. Furthermore, currents of of 1~nA and higher exhibit a weak connection between substrate and printed structure. The contrast has been modified for clarity. \textbf{(b)} Overlay of the outlines of different overpasses. The outline of each line was colored and is overlaid to illustrate the evolution of the structure over time. The difference between two overpasses therefore corresponds to the voxel that is deposited within 50ms. With currents of 0.25 and 0.5~nA, the structure grows over its entire width. For currents above 0.75~nA the growth for is confined to the highest section of the structure. \textbf{(c)} The shape of the voxel, however, does not only depend on the current, but also on the distance between nozzle and substrate. Here, a series of overlaid outlines is shown for lines that were printed with 0.75~nA and distances between 7.5\textendash 15 \textmu m. At a distance of 7.5 \textmu m again a porous structure was observed, probably due to dentritic growth inside a stationary ACN droplet. At a distance of 12.5 \textmu m the voxels start to become flattened out and at 15~\textmu m Cu is deposited onto a large area, probably due to a breakup of the ejected droplet. This leads to a spray of Cu ions instead of a confined electroplating. Therefore, the deposition with minimal feature size (\textit{i.e.} minimal thickness of the structure) requires an optimum between current and distance.}
  \label{fig_crosssection_ACN}
\end{figure*} 

\begin{figure*}[!ht]
 \centering
 \includegraphics[scale = 0.85]{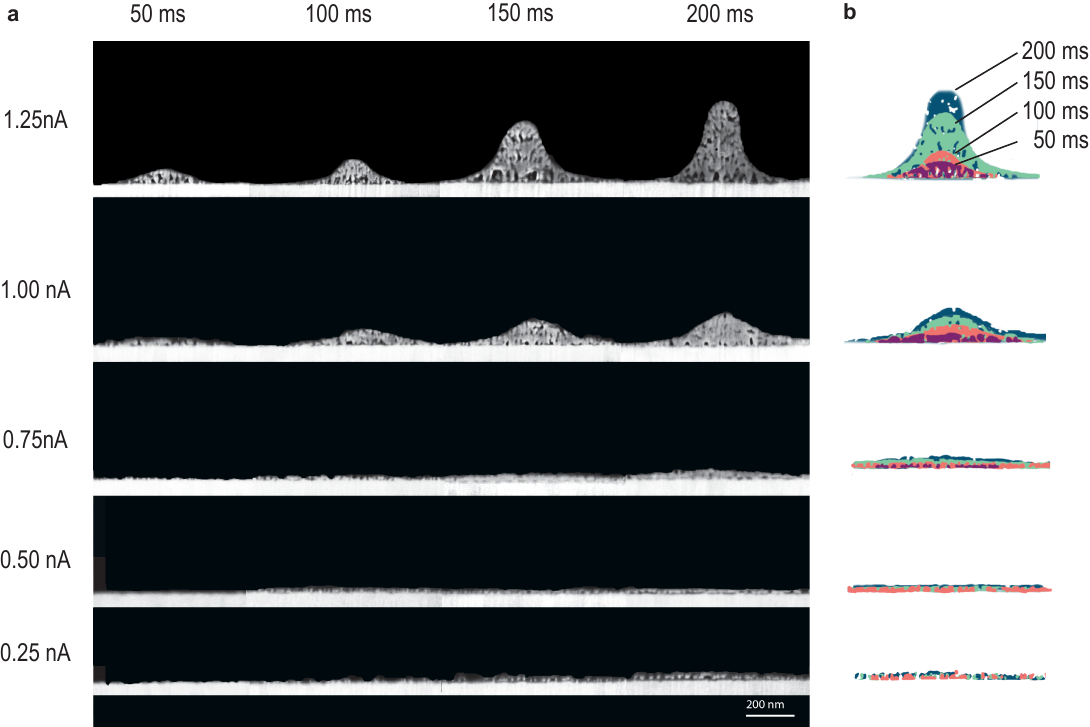}
 \caption{\textbf{Growth of Copper deposited from 1 mM CuSO\textsubscript{4} in H\textsubscript{2}O} \textbf{(a)} Crosssections of Cu lines printed with currents between 0.25\textendash 1.25~nA (nozzle-to-substrate distance of 7.5~\textmu m). The lines were printed with multiple overpasses of 50 ms each (for example, 150~ms refers to 3 overpasses). In contrast to ACN, flat and unconfined deposition is observed for currents $\leq$ 0.75~nA. Based on the observation in Fig. \ref{fig_crosssection_ACN}c, this indicates that the nozzle was too far from the substrate and the droplets evaporated in-flight. This leads to a spray and therefore unconfined deposition. Currents $\geq$ 1~nA show a a very fine microstructure with a high porosity.
 \textbf{(b)} Overlay of the outlines of the deposition show each additional voxel that was printed in 50~ms. Even for high currents, the deposition is not confined to the tip of the structure as for ACN, but growth over the entire width of the structure is observed.}
  \label{fig_crosssection_CuSO4}
\end{figure*} 

We observed that the current depends on a variety of parameters. During the second cycle between 0 and 160~V a higher current compared to the first cycle is measured for both acetonitrile and water as the solvent (Figure \ref{fig_CV} a,c). A third and forth cycle each lead to a further increase (SI \ref{fig_supp_Current}c). An increase in the nozzle-to-substrate distance, however leads to a decrease of the current and necessitates a higher V\textsubscript{ejection} to start the ejection. Water without an added electrolyte can be ejected with the same three different regimes of the conductivity (except for the highest nozzle-substrate distance, where the current stays too low), but requires high potentials to start the ejection. Consequently, an increased concentration of metal ions in the solvent lowers the potential needed to initiate ejection. The orifice size of the nozzle also has a pronounced influence on the ejection behaviour. While with a small aperture the above described behaviour is observed, a large nozzle orifice necessitates a large potential to start the ejection and the current increases steeply. Large orifices render the control over the current difficult due to this steep increase of the current.

Important for future depositions is the observation that the same applied potential can result in different currents depending on the elapsed time of the deposition and on parameters that are difficult to control precisely. For example, laser-based nozzle pulling systems produce orifices with a small variation in size and the distance between nozzle and substrate varies when out-of-plane structures are deposited. Therefore, the second part of this article investigates nucleation and growth dynamics under constant currents. To stabilize the current, the applied potential V is varied in such a way that the deposition current stays within $\pm$ 5~\% of the set value. 

\subsection{Nucleation and growth of Cu on Au}
Figs. \ref{fig_NucleationACN} and \ref{fig_NucleationCuSO4} show the nucleation of copper on gold substrates printed using ACN and 1~mM CuSO\textsubscript{4} dissolved in H\textsubscript{2}O, respectively, as solvent and electrolyte. The nucleation is shown for five different currents (0.25, 0.5, 0.75, 1 and 1.25~nA) and arrested at different time steps. These currents were selected to represent both sections of the CV graph that exhibit different conductivities. From previous studies it is also known that droplet-confined electroplating achieves fast deposition rates of around 3~\textmu m s\textsuperscript{-1}\cite{Menetrey2022TargetedPrinting}. Therefore, the necessary time steps to illustrate nucleation are in the order of few milliseconds. We found that the most reliable approach to such short time steps is to use a lateral translation of the substrate with different velocities (note that we keep the nozzle constant but move the substrate). The nominal residence time is then approximated by dividing the orifice diameter (ACN: 200\textendash  230~nm; H\textsubscript{2}O: 60\textendash  90~nm) by the velocity of the substrate (4\textendash 200 \textmu m s\textsuperscript{-1}). This approach was chosen as the reaction time of the control software and power source as well as parasitic capacities induced when switching on/off the system did not allow to reliably access such short time steps.

The nucleation behaviour of Copper in ACN (Fig. \ref{fig_NucleationACN}) shows distinct differences between low currents (0.2\textendash  0.5~nA) and high currents ($\geq$ 0.75~nA). The current and the necessary applied potential during deposition are shown in \ref{fig_supp_CVNucleation}. At 0.25~nA only patches of Cu are deposited. Number and size of the deposited patches increases with increasing residence time. At 0.5~nA very small grains are observed that  successively start to merge with longer residence times and form a rough, but continuous structure, as seen from SEM images and later crosssections. At currents of 0.75~nA and higher, progressively large grains can be observed when the residence time is larger than 20~ms. At 1~nA, few but very large grains can be observed, while at 1.25~nA dense lines without apparent grain boundaries can be observed. The area over which nucleation is observed (indicated by the dashed white lines when nuclei could be identified) depends strongly on the set current. At low currents, the width of the band stays constant, independent of the residence time. At high currents the nuclei are spread over a larger area (larger than the SE micrograph shows) and the area decreases with longer residence time. 

For copper that is deposited using water (Fig. \ref{fig_NucleationCuSO4}) generally smaller grains are observed than in ACN. Further, the width of the band, in which nuclei are observed stays constant over different currents. For high currents ($I \geq$ 1~nA) after a seeding stage, a few grains grow preferably before merging and forming a layer. This is shown with higher temporal resolution in Fig. \ref{fig_NucleationCuSO4}b. We identify three subsequent stages of the nucleation of Cu in water with 1.25~nA: first, very small nuclei are seeded homogeneously. In a second stage, the largest nuclei grow preferrably, until in a third stage, these grains start to coalesce and form a continuous line. Yet, a porous interface between Au substrate and the the Cu line remains. 

SI figure \ref{fig_supp_multipleoverpasses} shows the nucleation after the same number of time steps achieved by single or by multiple overpasses of a fraction of the residence time each, using 1~mM CuSO\textsubscript{4} as the electrolyte. The structures look similar after 10~ms residence time. The results indicate that for 1~nA using multiple, but shorter overpasses leads to an increase in the number of nuclei. Nevertheless, we conclude that using multiple short overpasses does not fundamentally change the growth behaviour and the mechanisms during the deposition.

\subsection{Illustrating the voxel during initial stages of the deposition}
The nucleation of Cu in different solvent and at various currents indicated strong differences in the resulting microstructure. These differences are shown in detail in Figs. \ref{fig_crosssection_ACN}a and \ref{fig_crosssection_CuSO4}a, that present the cross-sections of Cu lines printed with residence times of 50, 100, 150 and 200~ms at the indicated currents for both solvents. 50~ms residence time refers to a translational speed of 4 and 2 \textmu m/s with a nozzle with 200~nm and 100~nm orifice size respectively. At slower translational speeds, the lines are not adhering to the substrate anymore and out-of-plane growth is observed\cite{Galliker2012DirectNanodroplets}, because the growth-rate exceeds the velocity of the substrate. Therefore, multiple overpasses of 50~ms each are used to further increase the residence time without introducing out-of-plane growth. Further, this approach allows to illustrate the material deposited in 50~ms increments.

A current of 0.25~nA and ACN as the solvent deposits, as already shown in Fig. \ref{fig_NucleationACN}, only a small amount of material. This carbon-rich structure has a low contrast in SEM. However, it is still visible that this deposit shows a high porosity, resembling a network structure. At 0.50~nA a dense microstructure is observed. At 0.75~nA the deposition is confined to a narrower structure (after the first layer), meaning that material is predominantly added to the top section. Further, larger grains than in lower currents are observed. However, porosity starts also to emerge between the grains. A further increase of the current leads to wider (horizontal width) structures, larger grains and also more porosity in between the grains. In previous reports, a constriction of the base of printed pillars was reported\cite{Menetrey2022TargetedPrinting}. This is visible here as well for currents of 1~nA: It appears that the upper part of the structure is not connected to the substrate. This is probably due to the presence of porosity at the interface with a limited number of connecting grains between substrate and printed structure. 

In water, for currents of 0.75~nA and lower (and a nozzle-to-substrate distance of 7.5 \textmu m), a flat, unconfined deposition is observed (refered to as a spraying mode). Spraying occurs when the droplets reach the coulomb limit of the droplets leading to a spray of the droplet and hence a deposition in a larger area. Note that currents of 0.75~nA refer to the second regime in the I-V curves, before the slope changes. For currents of 1~nA and higher a very fine grainstructure is observed. This is in agreement with previous reports of Cu pillars printed with this technique \cite{Porenta2023Micron-scalePrinting} using 1~mM CuSo\textsubscript{4}. At 1.25~nA the structure exhibits high porosity, especially close to the interface with the substrate. This porosity was expected from Fig. \ref{fig_NucleationCuSO4} and similar porosity has been reported previously\cite{Nydegger2022AdditiveStructures}. 

The shape of the voxels, however, are a function of not only the current, but also the nozzle-to-substrate distance (Fig. \ref{fig_crosssection_ACN}c). An decrease in the distance leads to more porous structure, while at high distance spraying instead of confined deposition is observed. Therefore, optimal deposition parameter need to balance current and distance to achieve optimized material quality.

\section{Discussion}
A possible explanation of the observed current-potential characteristic as well as the differences in nucleation and microstructure of deposited Cu is the transition from a transient to a stationary electrolyte droplet on the substrate. A transient solvent droplet would evaporate before the next electrolyte droplet arrives. Such a transition has been hypothesized already in previous work based on an observed relationship between applied potential and microstructure\cite{Menetrey2022TargetedPrinting} for Cu and ACN, but not studied as a function of the current. 

At \textit{$V < V\textsubscript{ejection}$} no droplets are ejected and therefore no current is measured. Strikingly, the necessary potential to start any deposition is up to two order of magnitudes higher than in standard electroplating. This is due to the high electric fields necessary to drive the elecdrohydrodynamic ejection of the droplets. Further, the V\textsubscript{ejection} depends on the orifice size of the nozzle. For optimized deposition performance, an ideal orifice size leads to a linear I-V curve with a flat slope, which allows to finely control the current and does not lead to a deposition in a spray mode. This ideal orifice size was observed to differentiate between the utilized solvents. The ideal orifice size is probably unique for each solvent system as electrohydrodynamic ejection depends on a variety of parameters of the solvent\cite{Lee2013OptimizationPrinting} (such as the viscosity, surface tension, density, permittivity and conductivity). 

At \textit{$ V > V\textsubscript{ejection}\:\&\:I <$ 0.75 - 1~nA}, droplets are ejected but evaporate quickly, so that the solvent droplet evaporates completely before the next droplet reaches the substrate (or the droplets already evaporate before they impact the substrate), while at (\textit{$V > V\textsubscript{ejection}\:\&\: I > $0.8-1~nA:}), there is a stationary solvent droplet. The solvent droplet is partially screening the electric field between the substrate and the nozzle, requiring a higher applied potential for the ejection. This solvent droplet therefore acts as an additional resistance in the system, leading to a lower conductivity. Further, this hypothesis explains the observed changes in the nucleation between $I \leq$ 0.5~nA  and $I \geq$ 0.75~nA. Electroplating in a stationary droplet is less kinetically dominated and therefore leads to thermodynamically more favourable microstructures (\textit{i.e.} larger grains), as observed for high currents in ACN. In water, however, the trend towards larger grains is less visible. This could be due to higher levels of contamination (leading to frequent nucleation) or the formation of hydrogen on the substrate.

An alternative explanation for the observed different conductivities that we rejected is a change in the ejection mode\cite{Lee2013OptimizationPrinting}. There are different EHD modes described in the literature, which are, in order with ascending necessary electric field: dripping mode, pulsating cone, cone-jet, tilted-jet and twin- or multi-jet. A dripping mode is unlikely for our setup as this mode is generally associated with gravitational pull and the applied pressure to the solvent in the nozzle pushing, instead of an electrostatic pulling of the droplets. A tilted-jet is apparent in droplet-confined deposition by a deposition with varying offset from the nozzle position, while multi-jet deposition is apparent from multiple deposition sites simultaneously. Such an offset or multiple deposition locations were not observed for the discussed current-potential range. Hence, the two possible modes are pulsating-cone and cone-jet mode. Yet, a change in the mode would not explain the observed change in nucleation and growth. However, different ejection modes could explain the observed behaviour for too-large orifices (Fig. \ref{fig_CV}f): for a small nozzle, individual solvent droplets are ejected, while for a large nozzle only a jet can be sustained that connects nozzle and substrate (leading to the fast increase in current). A jet is most probably too small to be observed in the utilized camera.

The difference in the observed I-V characteristic between the first and the second cycle probably originates from the generation of a surplus of charge carriers during the first cycle (\textit{i.e.} more charge carrier generated than ejected, possibly due to the involved high potentials). This is further supported by the observation that the necessary V\textsubscript{ejection} is higher in pure water that in a solution of 1~mM CuSO\textsubscript{4}. Therefore and unsurprisingly a higher number of charge carriers increases the current at a given potential.

\section{Conclusions and Outlook}
Droplet-confined electroplating must be both predictable and controllable to be a versatile fabrication technique. Therefore, we investigated the current-potential relationship and explored how the current can serve as an \textit{in situ} feedback mechanism for the deposition. An analysis of the current unraveled a characteristic I-V curve. This curve allows to investigate and quantify the influence of a variety of parameters, such as the elapsed time, the distance between nozzle and substrate and the orifice size on the deposition. The deposition at constant currents allowed to catalogue the nucleation and growth of copper in both water and acetonitrile as an electrolyte.

For acetonitrile, a current of 0.75~nA indicates best shape of the voxel and a dense microstructure, although higher currents allow for lines without visible grain boundaries. An analysis of the current indicates that this is a transient regime for both studied electrolytes, where the droplets potentially still evaporate before the next droplet impacts. At current of 1~nA, the growth inside the solvent droplet leads to large (and sometimes very long grains). For water, the volumetric deposition speed is much lower at the same current, indicating a lower efficiency of the process, possibly due to other charge carriers, such as protons getting reduced on the substrate and forming hydrogen. 

The results presented here could also serve as a template protocol for testing novel candidate solvents for EHD-RP: first, a CV curve is measured to optimize the orifice size for the given solvent, similar to Fig. \ref{fig_CV}f. Then, lines are printed and cross-sections analyzed if material has been deposited. No deposition, but a measured ejection current can indicate no dissolution of the sacrificial anode \cite{Nydegger2022AdditiveStructures} or a decomposition of the electrolyte. If deposition in a spraying mode is observed as in Fig. \ref{fig_crosssection_CuSO4} for low currents, the distance should be reduced, while dendritic growth indicates too close a distance, too high a current (Fig. \ref{fig_crosssection_ACN}) or an insufficient vapor pressure of the solvent that prevents timely evaporation of the solvent. An iterative approach will allow to quickly optimize the deposition parameters (orifice size, set current and distance between nozzle and substrate) for a chosen solvent and metal combination. Such a study will provide insights into the interplay and importance of dielectric constant, surface tension and vapor pressure of the candidate solvent for successful droplet-confined electroplating.

In conclusion, the current offers a self-evident tool to predict the properties of the deposited copper. Since any slicing algorithm requires knowledge about the geometrical shape and volume of an individual voxel, utilizing the current as an indicator for the volumetric deposition rate is crucial for the controlled fabrication of intricate 3 dimensional shapes\cite{Menetrey2024NanodropletPrinting}. This readies the technique significantly to deposit functional structures.

\section{Experimental}
\subsection{Materials}
Nozzles for deposition were fabricated on a P-2000 micropipette puller system (Sutter Instruments) from filamented Quartz capillaries (Sutter Instruments, Item QF100-70-15). Nozzle diameters were determined on a Quanta 200F (Thermo Fisher Scientific), equipped with a Schottky type field emission gun (FEG) in low vacuum mode (30~Pa) to avoid charging. Nozzles with diameters of 60–90~nm were used for salt solutions and 200-230~nm for acetonitrile. The nozzles were filled with acetonitrile (Optima, Fisher Chemical)  or copper sulfate (1 mM CuSO\textsubscript{4} (Sigma Aldrich, 99.999\% metal basis) in water (LC/MS-Grade, Fisher Chemical)) by using gas-tight glass syringes with custom made PEEK tips. The nozzles were cleaned on the inside by rinsing with the solvent. Substrates were 0.4~cm × 2.0~cm pieces of Au coated (80~nm) Si wafers. The substrates were cleaned in technical acetone, analytical isopropanole and subsequently blow-dried with compressed air before use. Cu wires (Alfa Aesar, 0.25~mm diameter, 99.999\% metal basis) were etched in concentrated nitric acid (65\% HNO3, Sigma Aldrich) for 10~s, then dipped in water and stored in analytical ethanol until use (maximal 10 minutes), while Au wires (Metaux Precieux SA, 99.999\%) were immersed for 30~s, dipped in water and stored in air.

\subsection{Setup} 
Fundamentally, the setup consists of piezo stages that move the substrate in X, Y, and Z direction (QNPXY-500, QNP50Z-250, Ensemble QL controller, Aerotech). Additionally, stage translations in X and Y direction larger than 500~\textmu m were enabled by additional long-range stages (M112-1VG, PI for Y direction, manual micrometre screw, Mitutoyo for X direction). Piezo stages and power source were controlled through a custom Matlab script. The nozzle is mounted on a motorized nozzle holder (Z825B, controlled with Kinesis, both Thorlabs). The deposition is observed trough an optical microscope composed of a ×50 objective lens (LMPLFLN, Olympus) and a CMOS camera (DCC1545M, Thorlabs). The lens is mounted at an inclination of 60° to the substrate normal and the substrate is illuminated from the behind using a green LED (LEDMT1E, Thorlabs). A power source (B2962, Keysight) with triaxial cable connectors was used for polarizing the anodes. The metal wires were connected using a mechanical clamp. The complete printing setup is mounted inside a gas tight box to provide an oxygen-free atmosphere (Controlled with an oxygen sensor, Module ISM-3, Dansensor) and is placed on a damped SmartTable (Newport) to provide a vibration-free environment.

\subsection{Deposition}
Deposition was performed at room temperature in an argon atmosphere ($<$100~ppm O\textsubscript{2}). Typical potentials applied to the anode during printing were 80\textendash 160~V. For printing with a certain current the potential was adjusted to match this ejection current by averaging the current over 40 data points (spaced 0.2~s) and adjusted in 0.2~V increments if the average deviated more than 5\%. Nozzle substrate distance was controlled to be 5\textendash 15~\textmu m by analysing the light microscope images and adjusting the position of the z-axis piezo stage accordingly.
 
\subsection{Analysis} 
High Resolution Scanning electron microscopy (SEM) was performed with a Magellan 400 SEM (Thermo Fisher Scientific, former FEI) equipped with an Octane Super EDXsystem (EDAX, software: TEAM). Tilt angles were 55\textdegree. HR-SEM images were taken in immersion mode with an acceleration voltage of 5~kV. A dual-beam Helios 5UX (Thermo Fisher Scientific) with a focused Ga\textsuperscript{+} liquid metal ion source was used for focused ion beam (FIB) milling of  cross-sections. Prior to FIB-milling the pillar was coated by a protective carbon layer, which is visible in the SEM image as a black background. 

\section{Author Contributions}
M.N. and R.S. devised the concept. M.N. performed deposition experiments and provided SEM and FIB analysis. M.N. visualised the data and wrote the original paper draft. Both authors discussed the results and reviewed the manuscript.

\section{Conflicts of interest}
The Authors declare no conflicts of interest.

\section{Acknowledgement}
This work was funded by grant no. SNF 200021-188491. The authors thank Noa Bernasconi for initial deposition experiments. Electron-microscopy analysis was performed at ScopeM, the microscopy platform of ETH Zürich.

\bibliography{references} 
\bibliographystyle{rsc} 

\newpage
\newcommand{\beginsupplement}{%
        \setcounter{table}{0}
        \renewcommand{\thetable}{S\arabic{table}}%
        \setcounter{figure}{0}
        \renewcommand{\thefigure}{S\arabic{figure}}%
     }

\beginsupplement
\onecolumn
\noindent\Large{\textbf{Supplementary Information}}\\
\\
\\
\noindent\LARGE{\textbf{Nanoscale droplet-confined electroplating: nucleation, growth and coulombic efficiancy}} \\
\noindent\large
\\
Mirco Nydegger and Ralph Spolenak
\normalsize 

\vspace{2cm}
\setcounter{page}{1}

\begin{figure*}[ht]
 \centering
 \includegraphics{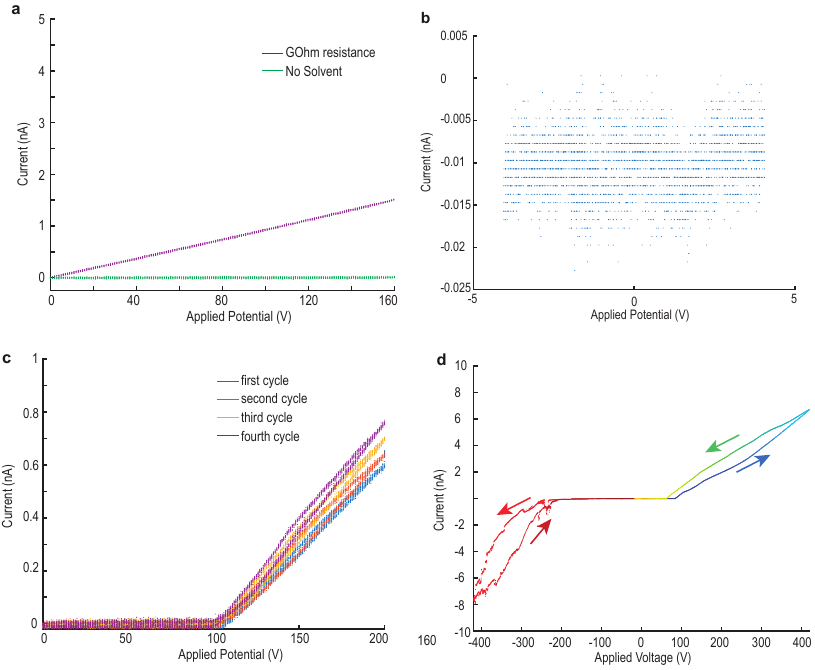}
 \caption{\textbf{: Current measurements} 
Here, a set of additional CV measurements are given to support the statements made in the article. \textbf{(a)} First, it is ruled out that the setup causes the observed potential-current behaviour. For example, a change in the internal resistance in the amperemeter could lead to a similar change in conductivity. Yet, when connecting a 100~GOhm resistor instead of the nozzle no change in the conductitivty is observed. Further, no parasitic currents could be observed when the potential is swept on a nozzle without solvent. \textbf{(b)} Potential-sweep around 0 did not indicate any current that could originate from a dissolution of the Cu anode in ACN (Cu Anode, ACN, 10 \textmu m distance, 50~mV/s). \textbf{(c)} Cycling multiple times between 0 and 160~V leads to an increase in the current for a given applied potential (above the minimal ejection potential). This could be driven by formation of surplus charge carrier in previous cycles or an activation of the surface, e.g. by a dissolution of a passivation oxide layer. \textbf{(d)} Full range potential sweep: The potential range of the utilized power supply is limited to $\pm$ 420~V. We swept this range from 0 to 420 to -420 to 0 Volts with 500~mV/s. Interestingly, the descending node of the current (after reaching 420~V, reducing towards -420~V) indicated a higher current than in the ascending node (rfom 0 to 420~V increasing). Again, this could be due to the generation of surplus charge carriers or an activation of the surface. Ejection with a negative potential (effectively inversing anode and cathode) is possible, yet it requires a higher minimal potential of -200~V to start the ejection. Further, the ejection is unstable and the current shows large variations. The origin of the variations is not known. SO far, no successful depositions have been carried out with EHD-RP utilising a negative applied potential.}
  \label{fig_supp_Current}
\end{figure*} 

\begin{figure*}[ht]
 \centering
 \includegraphics{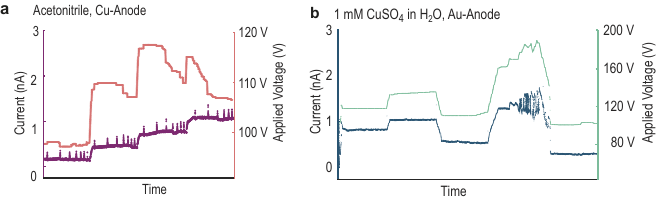}
 \caption{\textbf{: I-V curves during nucleation (a)} The current (left axis, purple) and potential (right axis, orange) curve during deposition of lines used to image the nucleation in acetonitrile with a sacrificial Cu anode (the curve for 1.25~nA is missing). The regular peaks of the current are induced by closing relais (connecting the anode to the power supply), as the wires were electrically isolated during regular distance calibrations. These calibrations ensured a correct distance between nozzle and substrate during the deposition. Interestingly, the necessary potential decreased during the deposition to ensure a constant current. This indicates an reduction of the resistance of the system. This could probably be caused by an easier dissolution of the sacrificial anode. \textbf{(b)} The current (left axis, blue) and potential (right axis, turquoise) curve during the deposition of the lines used to image the nucleation ,for 1~mM~CuSO\textsubscript{4} in H\textsubscript{2}O. In contrast to ACN, no reduction in the necessary potential to keep the current constant was observed. For deposition with an planned current of 1.25~nA, a large increase in the potential is observed. Further, the current is unstable exhibiting large variations.}
  \label{fig_supp_CVNucleation}
\end{figure*} 

\begin{figure*}[ht]
 \centering
 \includegraphics{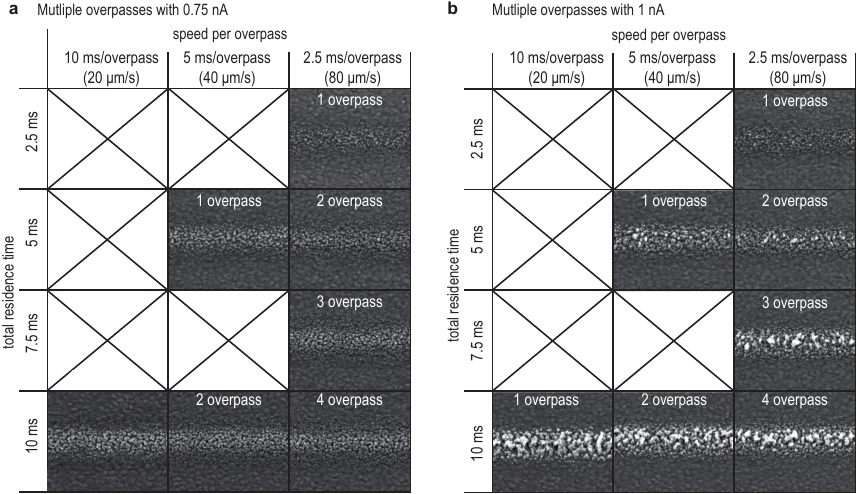}
 \caption{\textbf{: Difference in nucleation of Cu deposition from 1~mM~CuSO\textsubscript{4} between single vs multiple overpasses with equal total residence time} The nucleation is shown for a current of 0.75~nA in \textbf{(a)} and for 1~nA in \textbf{(b)}. The residence time is derived by dividing the orifice size of the nozzle by the velocity of the substrate during a translational movement. A total residence time of 10~ms is compared for 1 overpass, 2 (5~ms each) and 4 (2.5~ms each). for both currents no fundamental differences can be seen. It seems, however, that more and finer grains are observed for multiple overpasses. This could originate from new nuclei being formed in a second overpass and less time to form more stable larger grains.}
  \label{fig_supp_multipleoverpasses}
\end{figure*}

\end{document}